\documentclass{emulateapj}

\usepackage[]{natbib}

\newcommand{\kms}{{\rm km/s}}
\newcommand{\kpc}{{\rm kpc}}
\newcommand{\Rc}{r_{\rm c}}
\newcommand{\Vc}{V_{\rm c}}

\newcommand{\Msun}{M_\odot}
\newcommand{\mbh}{M_\bullet}
\newcommand{\ml}{{\rm M/L}}
\newcommand{\mlvobs}{{\rm M/L}_{V\rm,\;obs}}
\newcommand{\sigmar}{\sigma_r}
\newcommand{\sigmatang}{\sigma_t}
\newcommand{\sigmat}{\sigma_\theta}
\newcommand{\sigmap}{\sigma_\phi}

\shorttitle{The Supermassive Black Hole and DM Halo of NGC~4649}
  
\shortauthors{Shen \& Gebhardt}

\begin{document}
\title{The Supermassive Black Hole and Dark Matter Halo of NGC~4649 (M60)}

\author{Juntai Shen\altaffilmark{1} \& Karl Gebhardt}
\affil{Department of Astronomy, The University of
Texas at Austin, 1 University Station, C1400, Austin, TX 78712}
\altaffiltext{1}{Current address: Key Laboratory for Research in Galaxies and Cosmology, Shanghai Astronomical Observatory, Chinese Academy of Sciences, 80
  Nandan Road, Shanghai 200030, China}
\email{jshen@shao.ac.cn;gebhardt@astro.as.utexas.edu}

\slugcomment{Accepted for Publication in ApJ}

\begin{abstract}

We apply the axisymmetric orbit superposition modeling to estimate the
mass of the supermassive black hole and dark matter halo profile of
NGC~4649. We have included data sets from the Hubble Space Telescope,
stellar, and globular cluster observations. Our modeling gives $\mbh=
(4.5\pm 1.0) \times 10^9\Msun$ and $\mlvobs=8.7 \pm 1.0$ (or
$\ml_V=8.0\pm 0.9$ after foreground Galactic extinction is
corrected). We confirm the presence of a dark matter halo, but the
stellar mass dominates inside the effective radius. The parameters of
the dark halo are less constrained due to the sparse globular cluster
data at large radii. We find that in NGC~4649 the dynamical mass
profile from our modeling is consistently larger than that derived
from the X-ray data over most of the radial range by roughly 60\% to
80\%. It implies that either some forms of non-thermal pressure need
to be included, the assumed hydrostatic equilibrium may not be a good
approximation in the X-ray modelings of NGC~4649, or our assumptions
used in the dynamical models are biased. Our new $\mbh$ is about two
times larger than the previous published value; the earlier model did
not adequately sample the orbits required to match the large
tangential anisotropy in the galaxy center. If we assume that there is
no dark matter, the results on the black hole mass and $\mlvobs$ do
not change significantly, which we attribute to the inclusion of {\it
HST} spectra, the sparse globular cluster kinematics, and a diffuse
dark matter halo. Without the {\it HST} data, the significance of the
black hole detection is greatly reduced.

\end{abstract}


\section{Introduction} 
\label{sec:intro}

Most nearby galaxies harbor supermassive black holes at their
centers. Correlations between black hole (BH) mass and host galaxy
properties \citep{mag_etal_98,geb_etal_00,fer_mer_00,har_rix_04} have
been used extensively in theoretical models in order to understand
growth of the black hole and galaxy \citep[e.g.,][]{hop_etal_08}.  The
latest work from Hopkins et al. suggest that a main role of a black
hole is to halt star formation in the galaxy when the black hole is
large enough, thereby causing the dichotomy in colors
\citep[e.g.,][]{bell_08}. While there is still an active debate as to
the relative role of AGN feedback versus star formation feedback,
there is a consensus that physical mechanisms for black hole growth
are very important for understanding mass growth in galaxies. A
concern is that the black hole correlations may have significant
systematic biases, both from kinematics with poor spatial resolution
and models that do not adequately include the full mass profile (see
discussion in \citealt{gul_etal_09}).

Dynamical modeling of galaxies using orbit superposition offers one of
the best estimates on the black hole mass
\citep[e.g.,][]{rix_etal_97,van_etal_98,cre_etal_99,geb_etal_00,geb_etal_03,val_etal_04,tho_etal_04,tho_etal_05,sio_etal_09}. Assuming
axisymmetry, these models do not limit the form of the allowed
velocity anisotropies. Thus, the stellar orbital structure resulting
from the dynamical modeling provides a unique window on the mass
growth process in the massive systems, as long as the model
assumptions are valid.

A particular systematic bias is shown in \citet{geb_tho_09} where they
find that the black hole mass can be underestimated in the most
massive galaxies if the dark halo is not included. They find a
degeneracy between the dark halo and black hole mass, since without
the dark halo the stellar mass-to-light ratio is overestimated which
subsequently decreases the required contribution of the black hole to
constrain the central kinematics. In M87, \citet{geb_tho_09} find that
the black hole mass goes from $2.5\times10^9$ to $6.4\times10^9~\Msun$
by simply running models including a dark halo. Furthermore, the
uncertainties do not overlap, implying a large systematic effect. In
M87, however, this degeneracy is strong since the black hole is not
well resolved by the kinematic data. For galaxies with well
spatially-resolved kinematics, we do not expect the degeneracy to be
as significant.

In addition to studying the black hole mass, there is a strong need to
study the shapes of dark matter profiles. There is still little
consensus for those measured in individual galaxies (e.g., PNe from
\citealt{rom_etal_03}, stellar light from \citealt{for_geb_08}, X-rays
from \citealt{gas_etal_07,chu_etal_08,hum_etal_08,hum_etal_09}, and
globular clusters from \citealt{bri_etal_06,hwa_etal_08}). The
impressive work using gravitational lensing to measure the average
dark matter profiles (e.g., \citealt{man_etal_06a,man_etal_06b}) has
been able to reach out to nearly 1~Mpc. However, these results need to
be compared to measurements based on individual galaxies. It is
important to understand the galaxy-to-galaxy scatter in the profiles
and whether there are environmental effects.

This paper is part of an extensive campaign to measure both the black
hole mass and the dark matter profile simultaneously, to examine the
possible bias of the dark matter profile on the black hole mass
estimate for galaxies with various profiles, and to compare with the
gravitational potentials derived with other independent techniques
such as the X-rays and weak lensing studies. Initially, we focus on
the more massive galaxies, but it will be important to extend future
analysis over a large mass range. \citet{geb_tho_09} report results
for M87, the most massive nearby elliptical, and in this paper we
focus on NGC~4649. Both galaxies are giant ellipticals with a central
surface brightness ``core''.

In this paper we present the axisymmetric orbit superposition models
for NGC~4649 (M60), combining data from the Hubble Space Telescope
({\it HST}), stellar, and globular cluster observations.  NGC~4649 is a
giant elliptical with low surface brightness located in a subclump to
the east of the main Virgo concentration \citep{for_etal_04}.  NGC~4649
has been studied extensively for its total mass profile in recent
X-ray modelings \citep[e.g.,][]{hum_etal_06,gas_etal_07,hum_etal_08},
and globular cluster studies \citep[e.g.,][]{bri_etal_06}. The goals of
our study are to place NGC~4649's black hole mass estimate on a more
solid footing, to infer the properties of its dark matter halo, and
more importantly to offer an independent cross-check using different
dynamical tracers to the previous studies on NGC~4649.

We assume a distance to NGC~4649 of 15.7 Mpc. At this distance,
1$\arcsec$ corresponds to 76$\;$pc.


\section{Data}
\label{sec:data}

\begin{figure}[!ht]
\centerline{
\includegraphics[angle=-90.,width=0.9\hsize]{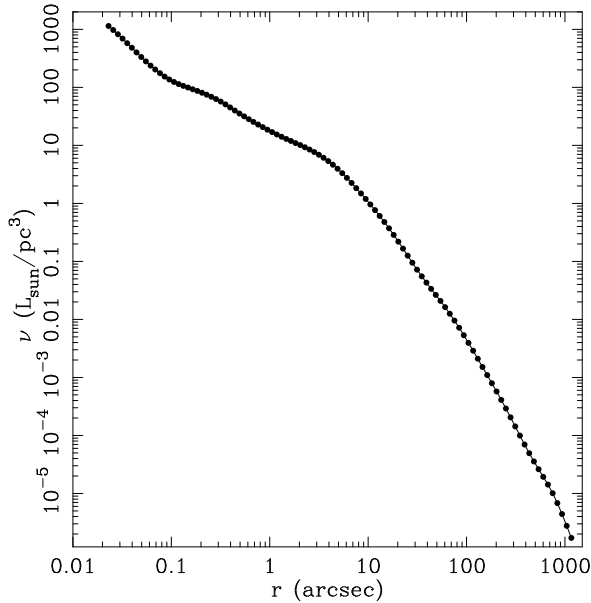}
}
\caption{The $V$-band volume luminosity density profile for stars. This comes from the deprojection of the surface brightness profile presented in \citet{kor_etal_09}.}
\label{fig:density}
\end{figure}

The input data of NGC~4649 (surface brightness profile and kinematics)
have all been previously published, and we rely on that data for the
models presented here. We use the stellar surface brightness profile
compiled by \citet{kor_etal_09} from a variety of sources, including
{\it HST} imaging as presented in \citet{lau_etal_05}.  All data from
\citet{kor_etal_09} have been transformed to $V$-band, and extend to
about 700\arcsec.  We deproject the surface brightness using Abel's
formula as in \citet{geb_etal_96}. The volume luminosity density we derived
is plotted in Figure~\ref{fig:density}.  The effective radius $R_e$ of
NGC~4649 is about $90 \arcsec$ by fitting a de Vaucouleurs profile to its
surface brightness profiles.

The stellar kinematics are the same as used in \citet{geb_etal_03},
and are presented in \citet{pin_etal_03}. These data include long-slit
observations on {\it HST} STIS and three position angles from
ground-based observations (see \citealt{pin_etal_03} for
details). There are no new stellar kinematic data presented in this
paper. The {\it HST} spectra extend to 0.8\arcsec, and the
ground-based stellar data extend to 70\arcsec. The dynamical modeling
code uses the line-of-sight velocity distribution (LOSVD), which are
the same as used in \citet{geb_etal_03}.

In order to extend to yet further radii, we also include data from
globular cluster (GC) velocities. The GC kinematics come from
\citet{hwa_etal_08}, which combine kinematic data from
\citet{bri_etal_06} and \citet{lee_etal_08}. There are 121 clusters
with velocities over a radial range from 32--533\arcsec. However, we
only use the velocities that are beyond 200\arcsec, since inside that
radius the clusters with velocities are spread over too many model
bins to add any significant information compared to the stellar
kinematics. A further complication with using GCs in the central parts
is that their number density profile is not well known there. Outside
of 200\arcsec, there are 61 clusters with velocities, and the average
radius of these clusters is 319\arcsec \ \citep{hwa_etal_08}. Rotation
in the GC system at radii beyond 200\arcsec\ is reported in
\citet{hwa_etal_08}; however, \citet{bri_etal_06} argue for little to
no rotation at radii around 200\arcsec. We therefore explore models
where we include the rotation directly (by fitting a velocity profile
on the major axis at 319\arcsec\ with a rotation amplitude and
velocity dispersion as reported by \citealt{hwa_etal_08}) and where we
use the second moment of the GC velocities  only (by fitting one
velocity profile, centered on zero velocity, to the GC velocities in
the full annulus). The results are nearly identical.  In this paper,
we only present results from using the second moment. Our
analysis of the Hwang et al. data implies a second moment of
267($\pm25$)~\kms, which is consistent with the analysis of \citet{bri_etal_06}. Since the modeling code uses LOSVDs directly, we must
transform the moments into LOSVDs, which we do using Monte Carlo
realizations of the values and uncertainties (as discussed in
\citealt{geb_etal_03}).

The dynamical models require the number density profile of the GCs
(since we use clusters as a tracer population). \citet{hwa_etal_08}
present number density profiles of the GCs. We use their equation 5
for the combined sample as the deprojected number density
profile. Comparing this profile to the stellar luminosity density
shows that beyond 150\arcsec, the two profiles have the same
slope. Inside that radius, the GC number density profile flattens more
compared to the stars. Since we do not use any clusters inside of
200\arcsec, we cadopt the same number density profile for
the clusters as we do for the stars.

\begin{figure*}[!ht]
\centerline{
\includegraphics[angle=-90.,width=0.75\hsize]{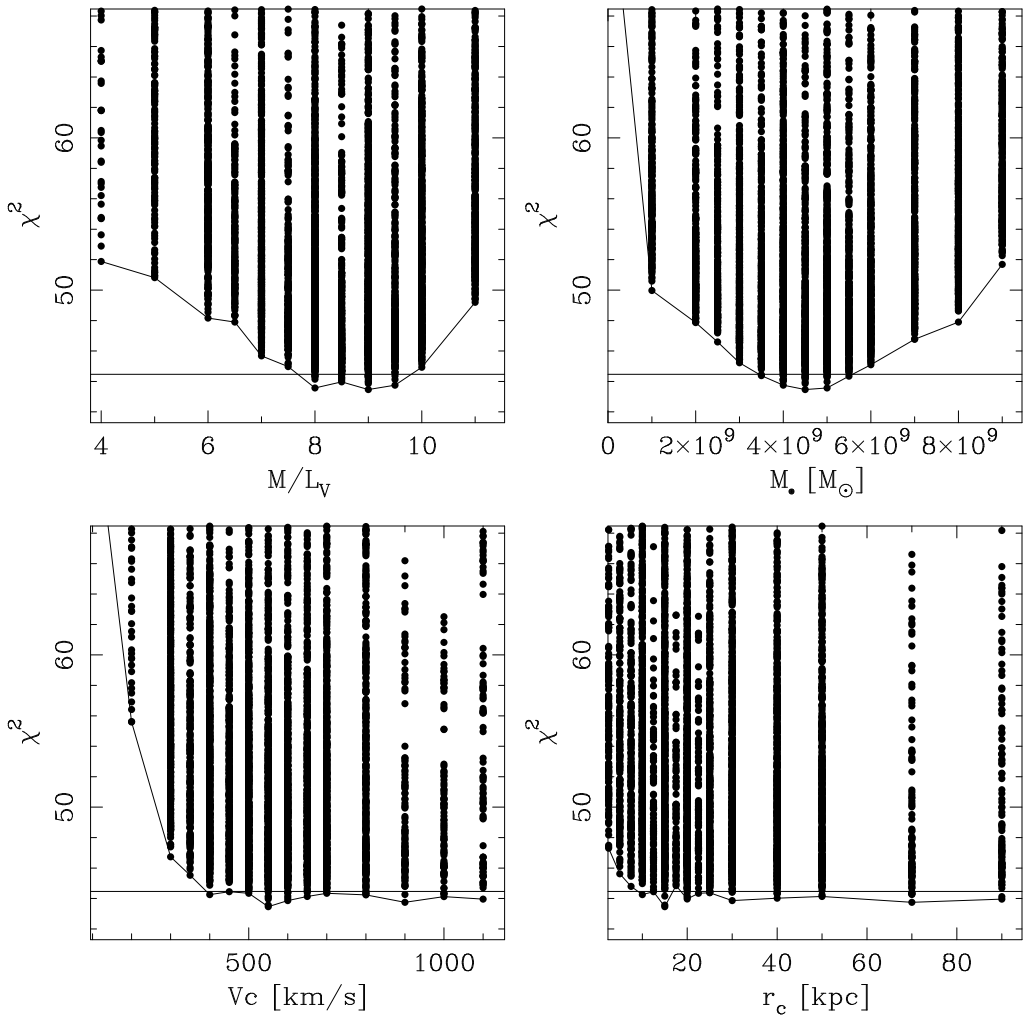}}

\caption{$\chi^2$ versus $\mlvobs$, black hole mass $\mbh$, the
halo scale velocity $\Vc$, and the halo core radius $\Rc$. Each point
represents a possible model for the logarithmic halo (only models with
$\Delta \chi^2\la 25$ over the minimum are shown in order to highlight
the $1\sigma$ uncertainty). The solid line along the bottom ridge
represents the marginalized $\chi^2$ values which we use to determine
the best fit and uncertainties. The horizontal line marks the 68\%
confidence limit ($\Delta \chi^2=1.0$).}
\label{fig:chi2_4p}
\end{figure*}

\begin{figure*}[!ht]
\centerline{
\includegraphics[angle=-90.,width=0.75\hsize]{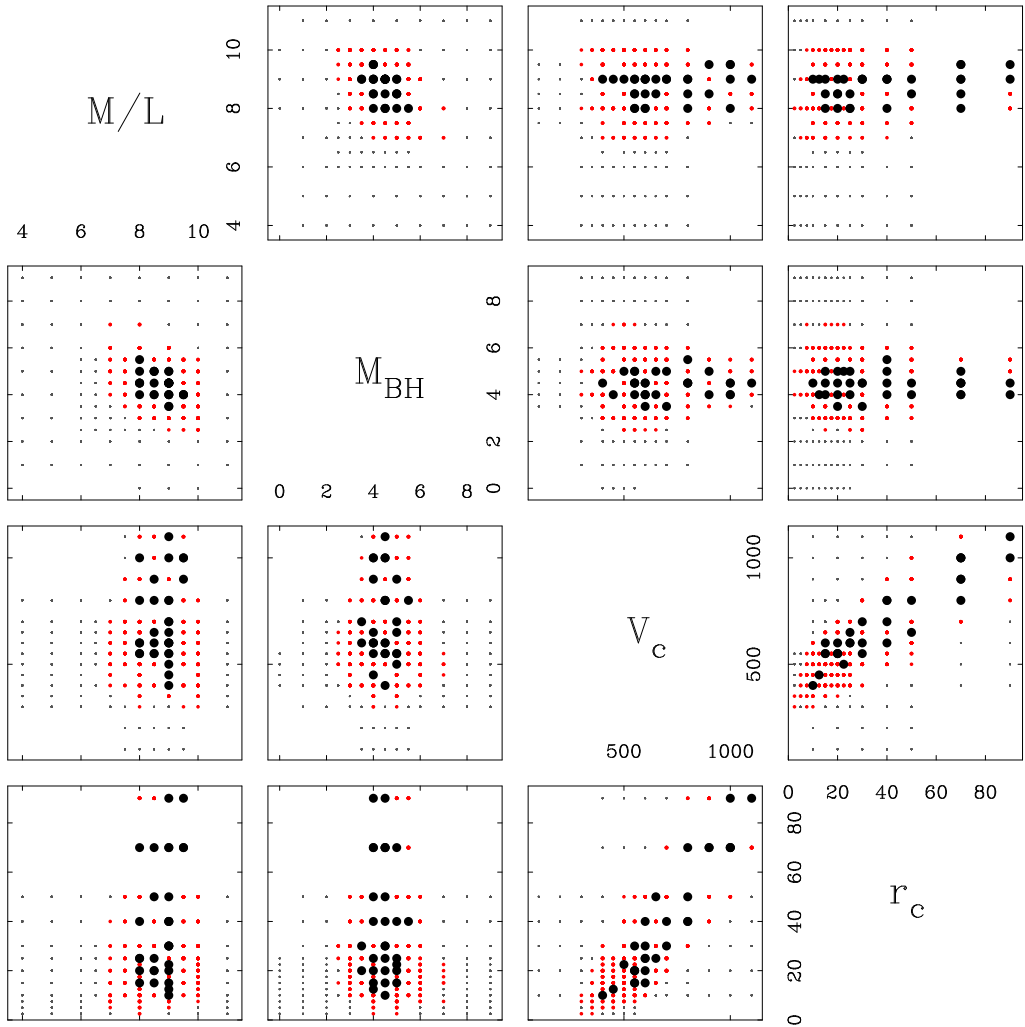}}

\caption{Plots of the $V$-band $\mlvobs$, black hole mass $\mbh$ (in units
of $10^9\Msun$), the scale velocity $\Vc$ of the logarithmic halo in
$\kms$, and the core radius $\Rc$ of the logarithmic halo in kpc
against each other to show possible degeneracies. The small grey points
represent the locations of all the models. The large black points are
the locations of those models that are within the 68\% confidence
limit ($\Delta \chi^2=1.0$) after marginalizing over the other
possible parameters. The red points are those that are within the 95\%
confidence band. The strong correlation between $\Vc$ and $\Rc$ is
apparent.}
\label{fig:corr}
\end{figure*}

\section{Dynamical Models}
\label{sec:models}

The orbit superposition models we use are based on the idea invented
by \citet{schwar_79}, and are described in detail in
\citet{geb_etal_00,geb_etal_03,tho_etal_04,
tho_etal_05,sio_etal_09}. The models have been tested extensively to recover both
the black hole mass and the dark matter halo without significant bias
for axisymmetric systems.

We first determine the luminosity density from the surface brightness
profile as described in \S\ref{sec:data}. Although we have
constructed models using different inclinations, we generally assume
that the galaxy is edge-on and adopt the minor-to-major axial ratio of
0.9. Next we compute the gravitation field from the 3-D light
distribution assuming that the mass consists of a black hole with mass
$\mbh$, and stars with a constant $\mlvobs$ ratio independent of position,
and a dark matter halo (now parametrized with only logarithmic and NFW
profiles).

Next our orbit-based model is implemented in two steps: constructing
the orbit library and fitting the full LOSVD (as opposed to using only
the second velocity moments) with the orbit library.  We measure the
LOSVD in radius and angle relative to the symmetry axis of the galaxy,
and in velocity. We use 24 radial, 5 angular, and 15 velocity
bins. The radial range of the dynamical models extend to
1000\arcsec. We calculate the galaxy potential and force on a grid
that is 5 times finer than the grid used in the data comparison. We
carefully choose the initial positions of the orbits to generate a
sufficiently dense sampling of $E$, $L_z$, and $I_3$
\citep{tho_etal_04}. This ensures a reliable representation of the
full phase space. Generally we have at least $30,000$ orbits when both
the black hole and dark matter halo are included.

We then choose the orbital weights so the superposition of orbits
match the light distribution and LOSVDs of the galaxy as well as
possible. The orbital weights are derived from the maximization of the
objective function $\hat{S}=S-\alpha\chi^2$ where $S$ is an
approximation to the usual Boltzmann entropy and $\chi^2$ is the sum
of squared residuals to the data. The smoothing parameter $\alpha$
controls the influence of the entropy $S$ on the orbital weights. We
cannot specify the optimal value of $\alpha$ in advance. In practice
we start with a very small $\alpha$ and gradually increase it until
the improvement in $\chi^2$ is less than a given percentage in a single
iteration. We have verified that our results are not sensitive to
the choice of the smoothing parameter $\alpha$.

We ran nearly 16,000 different models to estimate the black hole
mass and the dark matter halo parameters. Each model has a distinct
set of orbit library, and takes about 1.5 hours on the supercomputer
{\it lonestar} of the Texas Advanced Computing Center.


\section{Results}
\label{sec:results}

\subsection{Models including a dark matter halo}
\label{sec:canonical}
In our canonical set of models, the mass distribution $\rho$ of
NGC~4649 consists of a central black hole, stellar mass density, and a
dark matter halo:

$$\rho = \mbh \delta(r) + \Upsilon\nu + \rho_{\rm DM}$$

where $\Upsilon=\mlvobs$ is the mass-to-light ratio of the stars (a
position-independent constant), and $\nu$ is the observed $V$-band
stellar luminosity density. $\mlvobs$ needs to be corrected with
the foreground Galactic extinction in order to compare with that
derived from stellar population models. A popular foreground Galactic
extinction value $A_V=0.088$ \citep{sch_etal_98} could be adopted; it
gives the extinction-corrected $\ml_V =  \mlvobs /1.084$.

We describe the dark matter halo with a logarithmic profile, whose density is given as
$$\rho_{\rm  DM} = \frac{\Vc^2}{4\pi G}\frac{3\Rc^2+r^2}{(\Rc^2+r^2)^2}$$
and potential as
$$\Phi=\frac{1}{2} \Vc^2 \ln (1+\frac{r^2}{\Rc^2}).$$

\citet{geb_tho_09} found that realistic NFW models that fit the
globular cluster kinematics are not centrally concentrated enough to
dominate the mass in the inner regions; the enclosed mass profile they
get with NFW models are very similar to those obtained with
logarithmic halos. Thus, whether we parameterize with an NFW or a
logarithmic halo will have little influence on the black hole mass
estimate, particularly true for NGC~4649 where HST spectra exist. We
focus on a logarithmic halo in this study.

The kinematics include all data sets from the {\it HST}, stellar, and
globular cluster observations (\S\ref{sec:data}).  We use the $\chi^2$
distribution from all possible models to determine the four best-fit
parameters ($\mlvobs$, $\mbh$, $\Vc$, $\Rc$), and the associated
uncertainties. We start with a uniform but sparse grid in the 4-D
parameter space, then sample the smallest $\chi^2$ region with a finer
grid. As we will see from the figures below, the $\chi^2$ minimum and
the contours of NGC~4649 are quite regular, so our sampling procedure
should be adequate to cover the parameter space near the best-fit
values. The uncertainties in the parameters are determined from the
change in the marginalized $\chi^2$ as we vary one of the variables;
$\Delta \chi^2=1$ above the minimum represents the 68\% confidence
band or 1$\sigma$ uncertainty. We use the middle of the 68\%
confidence range as the best-fit value and half of that range as
the uncertainty.

\begin{figure}[!ht]
\centerline{
\includegraphics[angle=-90.,width=\hsize]{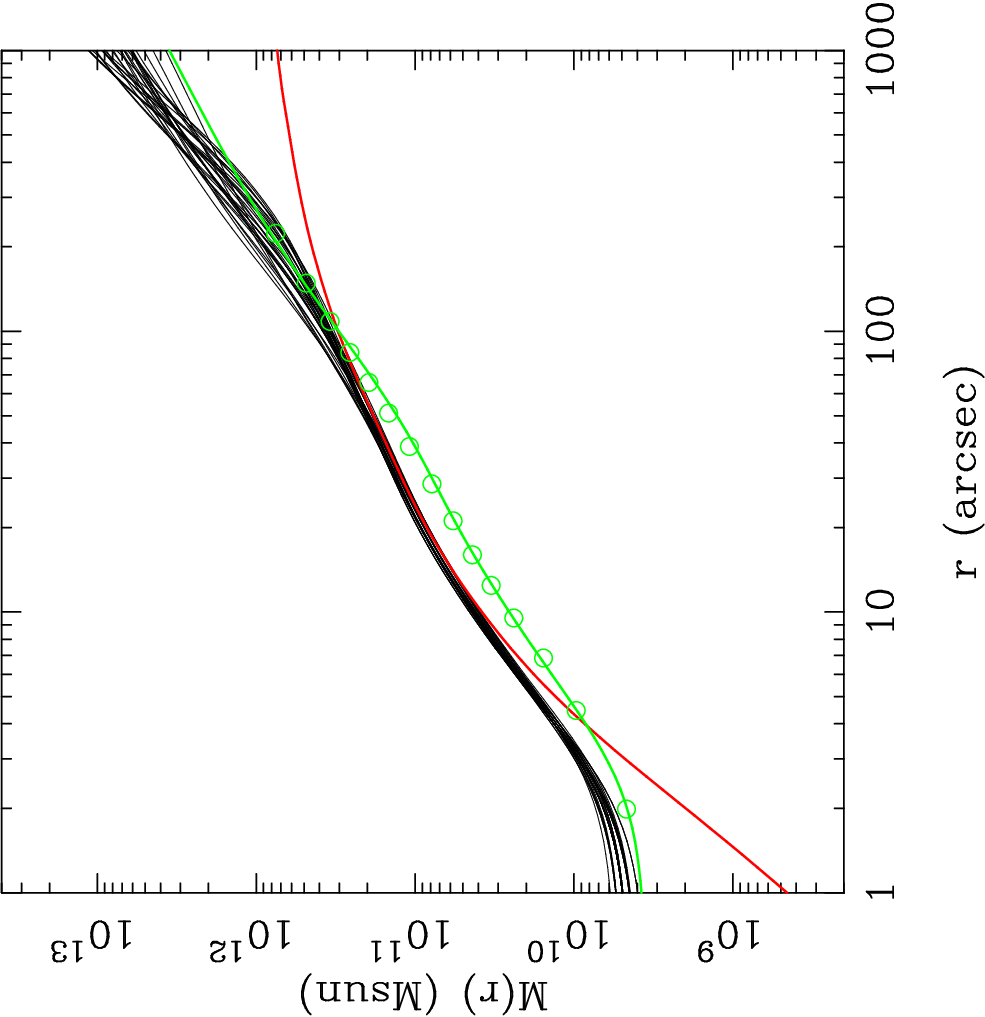}}

\caption{The mass profile for NGC~4649. The black lines represent the
models that are within the 68\% confidence band of the best fit (as in
Figure~\ref{fig:corr}). The green line is our representation of the
X-ray derived enclosed mass profile (the green circles, as in
\citealt{gas_etal_07,hum_etal_08}). The parameters for the green line
are $\mbh=3.5\times 10^9 \; \Msun$, $\mlvobs=5.0$, $\Vc=410 \kms$, and $\Rc
=10 \kpc$. The red line is the average contribution from the stars,
where we integrate the light profile in Figure~\ref{fig:density} then
times 8.7 (the best-fit $\mlvobs$). Clearly the total mass profile
obtained from our modeling is consistently larger than that from
X-rays over most of the radial range. }
\label{fig:dm}
\end{figure}

\begin{figure}[!ht]
\centerline{
\includegraphics[angle=-90.,width=0.8\hsize]{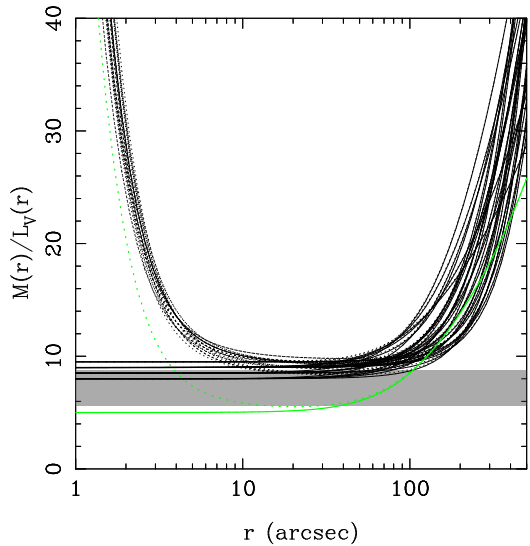}}

\caption{The integrated $\mlvobs$ as a function of radius.  The black
dotted lines represent the models that are within the 68\% confidence
band of the best-fit potentials (black lines in 
Figure~\ref{fig:dm}). The black solid lines are the same models but
excluding the contribution from the black hole. The green solid and
dotted are our best representation of the X-ray potential without and
with a black hole.  The shaded horizontal band is the range of
possible $\mlvobs$ (after applying the foreground Galactic extinction
of $A_V=0.088$) from stellar population models \citep{marast_98,
marast_05}, depending on what form of IMF is assumed.  }

\label{fig:ML_r}
\end{figure}

Figure~\ref{fig:chi2_4p} presents $\chi^2$ as a function of $\mlvobs$,
the black hole mass $\mbh$, the halo scale velocity $\Vc$, and the
halo core radius $\Rc$, including all values for the other three
parameters. Each point represents a possible model, but we show only
models near the $\chi^2$ minimum to highlight the $1\sigma$
uncertainty. The solid line along the bottom ridge represents the
marginalized $\chi^2$ values which we use to determine the best fit
and uncertainties. Given the number of LOSVDs, the velocity binning
used in the modeling, and that the LOSVD bins are correlated (as
discussed in \citealt{geb_etal_03}), the reduced $\chi^2$ is around
0.4, which is typical with orbit-based models.

From Figure~\ref{fig:chi2_4p}, we find the black hole mass $\mbh=
(4.5\pm 1.0) \times 10^9\Msun$ and the stellar $\mlvobs=8.7 \pm 1.0$
(or $\ml_V=8.0\pm 0.9$ after foreground Galactic extinction is
corrected with $A_V=0.088$). The scale velocity $\Vc$ and the core
radius $\Rc$ of the logarithmic halo are not well constrained. As we
will discuss later, halo parameters are degenerate with each other,
and their exact values have little effect on the shape of the dark
matter profile in the region of interest.

In order to find the possible correlations and degeneracies in the
four parameters, Figure~\ref{fig:corr} plots $\mlvobs$, $\mbh$, $\Vc$
and $\Rc$ against each other. The small grey points represent the
locations of all the models. The large black and red points highlight
those models that are within the 68\% ($\Delta \chi^2=1.0$) and the
95\% ($\Delta \chi^2=4.0$) confidence bands, respectively, after
marginalizing over the other possible parameters. The strong
correlation between $\Vc$ and $\Rc$ is apparent. This is because the
dynamical models only fit for the enclosed mass and we do not have
enough kinematics at large radii; we cannot distinguish dark halos
with a large $\Vc$ and a large $\Rc$ from the one with a small $\Vc$
and an accordingly fine-tuned small $\Rc$. By the same token, the
degeneracy between $\mlvobs$ and $\mbh$ (apparent in the the 95\%
confidence band) is expected; the decreasing/increasing contribution
from the stars can be made up by having a larger/smaller black hole
mass, to certain extent.  For NGC~4649, the degeneracy between
$\mlvobs$ and $\mbh$ is relatively weak because the inclusion of {\it
HST} data makes the uncertainties in $\mbh$ much smaller, unlike in
M87's case \citep{geb_tho_09}. {\it HST} spectra provide good spatial
sampling inside of the region where the black hole dominates over
stars. We do not find obvious degeneracies among other parameters.

The radial profile of the enclosed mass in NGC~4649 is shown in
Figure~\ref{fig:dm}. The black lines represent the models that are
within the 68\% confidence band of the best fit (large black points in
Figure~\ref{fig:corr}). The red line stands for the stellar
contribution alone using the best-fit $\mlvobs$ of 8.7. The stellar
component clearly dominates the mass profile from $10\arcsec$ to
$100\arcsec$. At the effective radius $R_e$ ($\sim$ 90\arcsec), the
best-fit stellar mass accounts for about 75\% of the total enclosed
mass. Within $10\arcsec$ the $4.5 \times 10^9\Msun$ black hole is
dominant, while outside $100\arcsec$ the dark matter halo
prevails. Compared with NGC~4649, the best-fit dark halo of M87 is
much more dominant \citep{geb_tho_09}. This also shows that
ellipticals with comparable luminosities can have quite different
distributions of dark matter relative to stars.

Figure~\ref{fig:ML_r} presents the integrated mass-to-light ratio
$\mlvobs$ as a function of radius. The black dotted lines are the
total $\mlvobs$ of the models that are within the 68\% confidence band
of the best-fit potentials (black lines in Figure~\ref{fig:dm}). The
black solid lines are the same models but excluding the contribution
from the black hole.  Consistent with Figure~\ref{fig:dm},
Figure~\ref{fig:ML_r} shows that the increase of the total $\mlvobs$
at small radius is dictated by the black hole, and the increase at
large radius is due to the dark matter halo. In the intermediate
radial range, from $10\arcsec$ to $100\arcsec$, the $\mlvobs$ is a
constant that stellar contribution dominates.

\begin{figure}[!ht]
\centerline{
\includegraphics[angle=-90.,width=0.7\hsize]{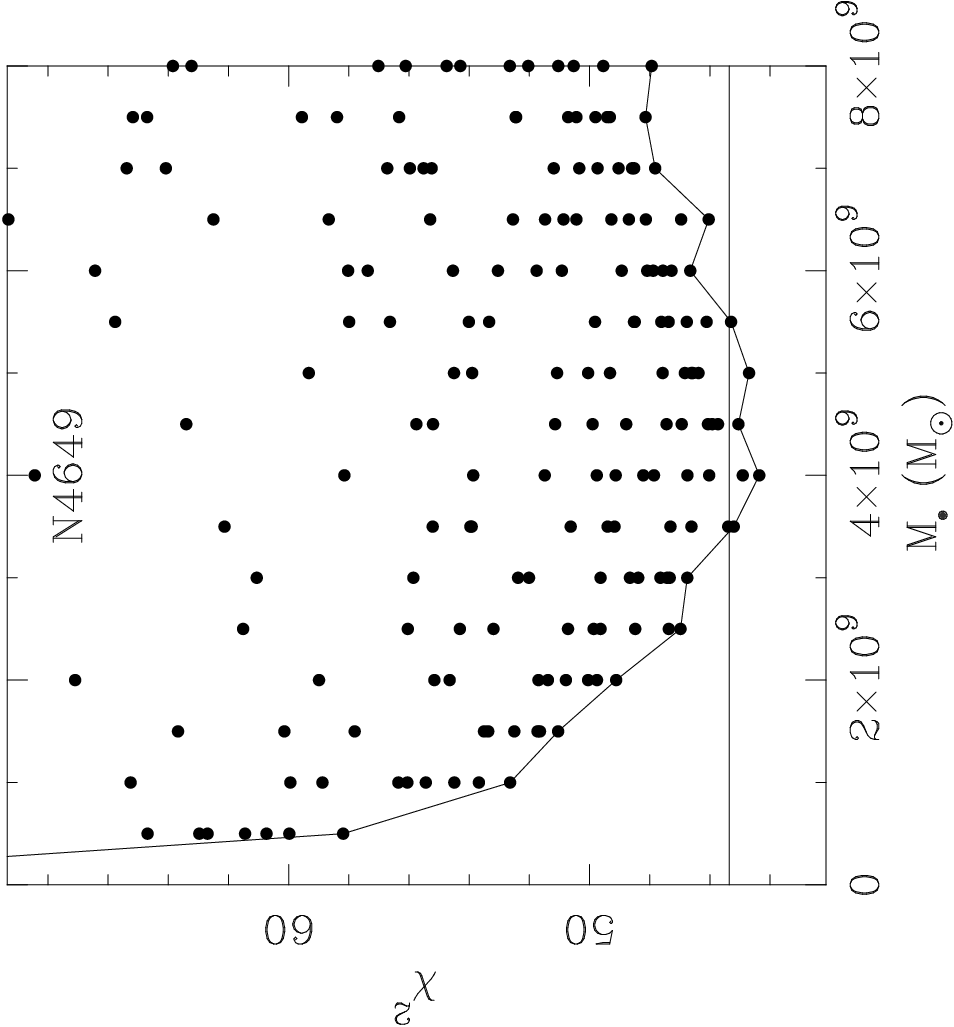}}
\vspace{0.02\hsize}
\centerline{
\includegraphics[angle=-90.,width=0.7\hsize]{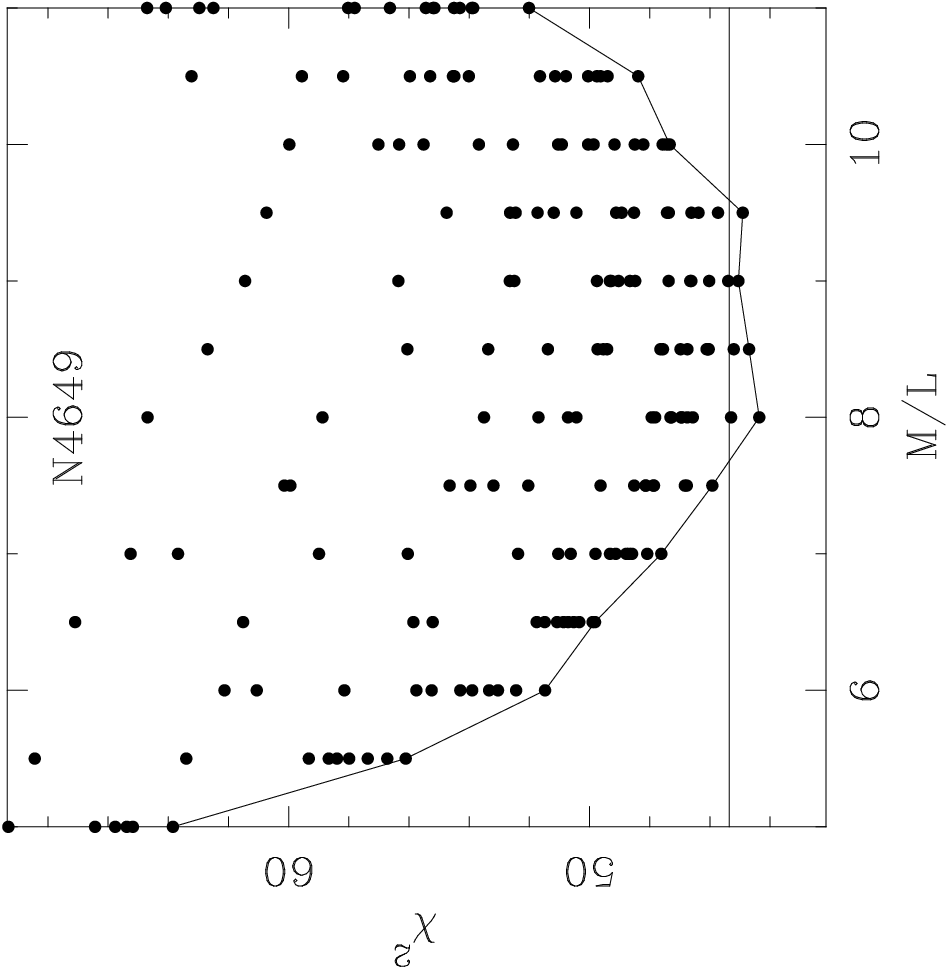}}
\vspace{0.02\hsize}
\centerline{
\includegraphics[angle=-90.,width=0.7\hsize]{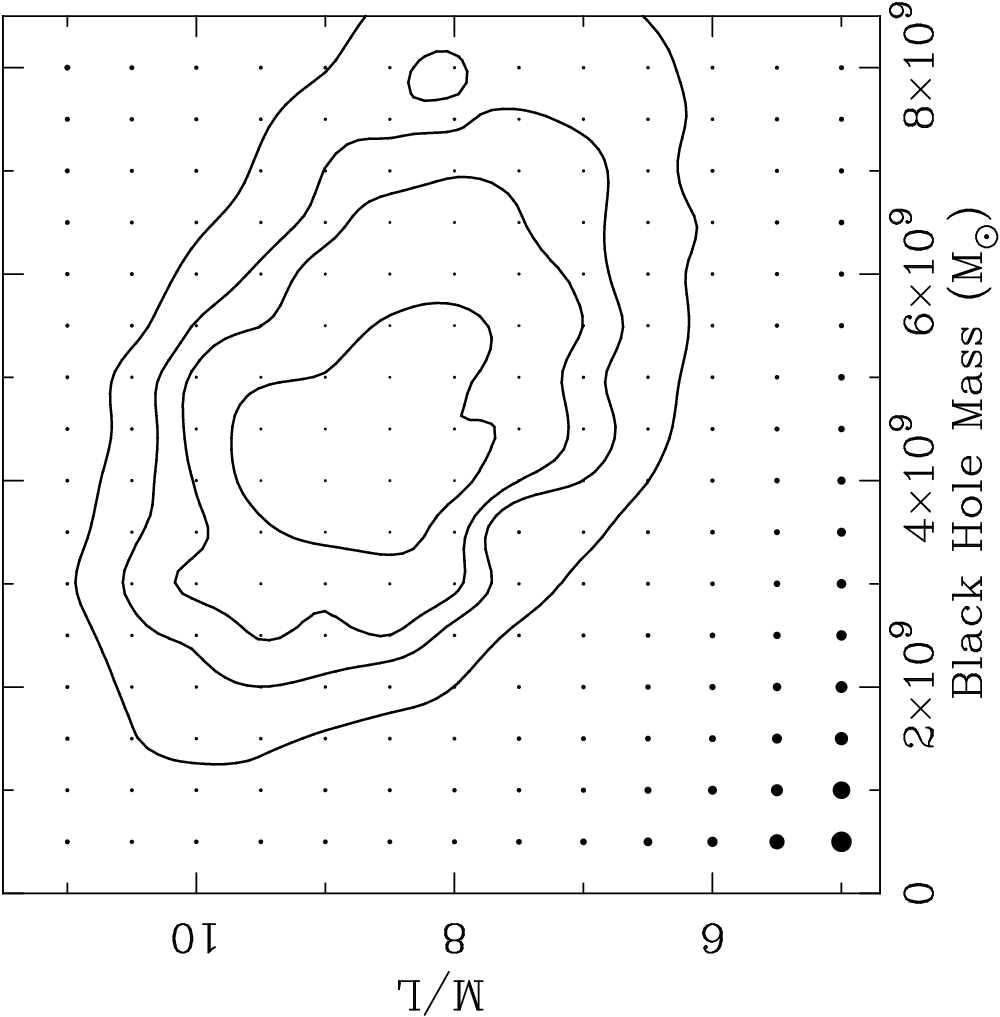}}

\caption{$\chi^2$ distribution when a logarithmic halo with $\Vc=410
\kms$ and $\Rc =10 \kpc$, which best matches the results of X-ray
modeling (the green line in Figure~\ref{fig:dm}), is imposed in the
modeling. (a): $\chi^2$ as a function of black hole mass $\mbh$,
including all values in $\mlvobs$. Each point represents a model, however,
not all models that we ran are shown in order to highlight the
$1\sigma$ uncertainty. The solid line along the bottom ridge
represents the marginalized $\chi^2$ values which we use to determine
the best fit and uncertainties. The horizontal line marks the 68\%
confidence limit ($\Delta \chi^2=1.0$). (b): as in (a) but for $\chi^2$
as a function of $V$-band $\mlvobs$, including all values in $\mbh$. (c): 2-D
plot of $\chi^2$ as a function of black hole mass and $\mlvobs$. The
points represent models that we ran. As in \citet{geb_etal_02}, the
contours were determined by a 2-D smoothing spline interpolated from
the these models and represent $\Delta \chi^2$ of 1.0, 2.71, 4.0 and
6.63 (corresponding to 68\%, 90\%, 95\%, and 99\% for 1 degree of
freedom).}
\label{fig:xrayfit}
\end{figure}

\begin{figure}[!ht]
\centerline{
\includegraphics[angle=-90.,width=0.9\hsize]{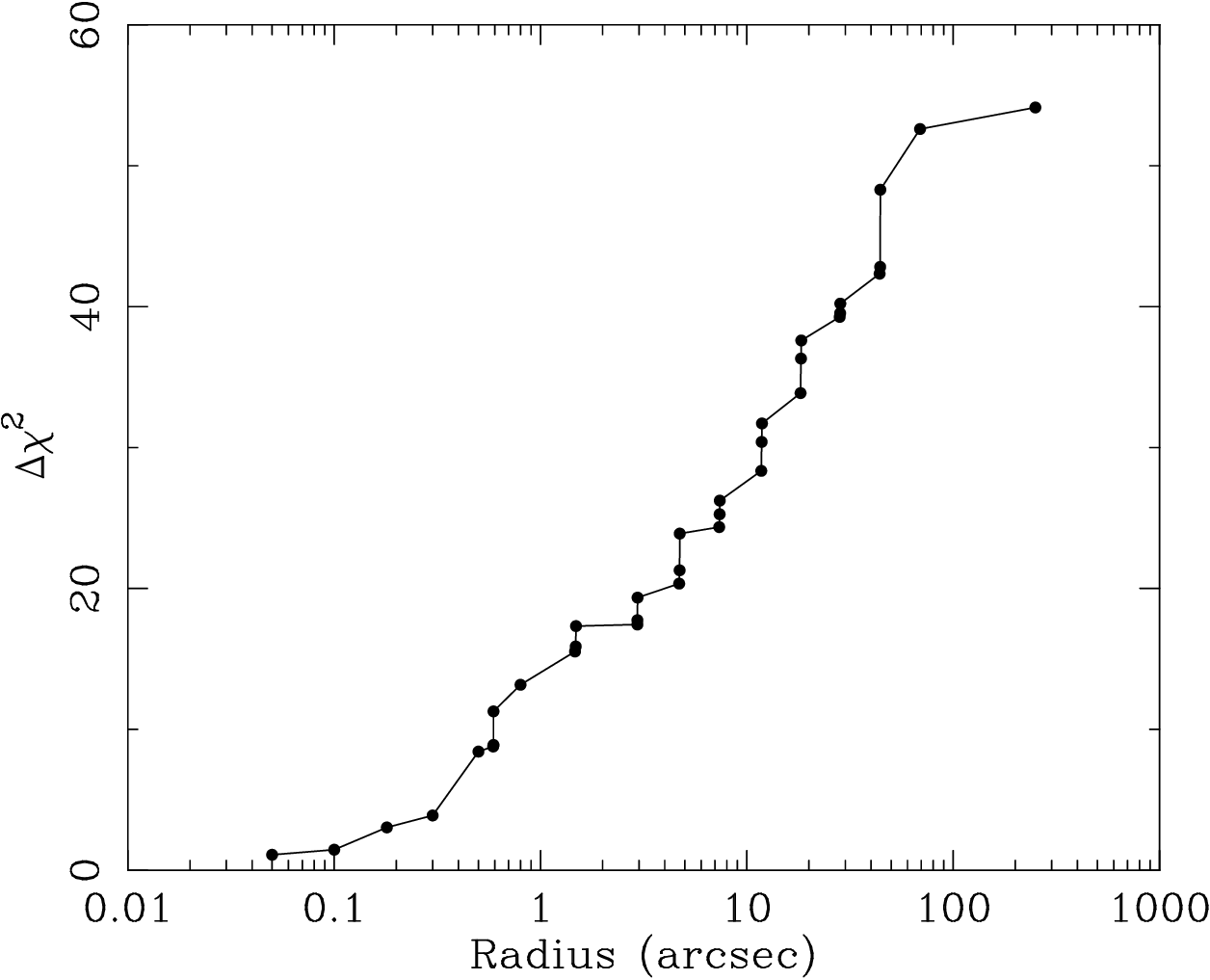}}

\caption{The radial profile of $\Delta \chi^2$, which is the
difference in the radius-cumulated $\chi^2$ between the best-fit
model ($\mbh=4.5\times 10^9 \; \Msun$, $\mlvobs=9.0$, $\Vc=550 \kms$, and
$\Rc =15 \kpc$) and the model that best matches the X-ray derived
potential ($\mbh=3.5\times 10^9 \; \Msun$, $\mlvobs=5.0$, $\Vc=410 \kms$,
and $\Rc =10 \kpc$).}
\label{fig:dchi2_r}
\end{figure}

The shaded band in Figure~\ref{fig:ML_r} shows the wide range of
possible $\mlvobs$ from stellar population models derived as below. We
first adopt the age ($\approx$ 12 Gyr) and metallicity $[Z/H]\approx
0.25$ for NGC~4649 from \citet{tra_etal_00}.
With the age and metallicity, we then interpolate for the stellar
$\ml_V$ from
tables\footnote{{http://www.icg.port.ac.uk/\textasciitilde
maraston/SSPn/ml/ml\_SSP.tab}} produced by the evolutionary population
synthesis models \citep{marast_98, marast_05}. If the Salpeter IMF
(0.1 to 100 $\Msun$) is assumed, then we get the interpolated stellar
$\ml_V=8.06$ as the upper bound of the shaded area. If the Kroupa IMF
(0.1 to 100 $\Msun$) is used instead, then we obtain the interpolated
$\ml_V= 5.13$ as the lower bound. In order to compare with our
uncorrected $\mlvobs$ in Figure~\ref{fig:ML_r}, we have to multiply
the $\ml_V$ from stellar population models by a factor of 1.084
($A_V=0.088$) to mimic the foreground Galactic extinction. From
Figure~\ref{fig:ML_r} we see that our best fit $\mlvobs=8.7\pm 1.0$
agrees best with the stellar population modeling result assuming the
Salpeter IMF, but it is also consistent with the range of results
assuming different IMFs.

\begin{figure}[!ht]
\centerline{
\includegraphics[angle=-90.,width=0.7\hsize]{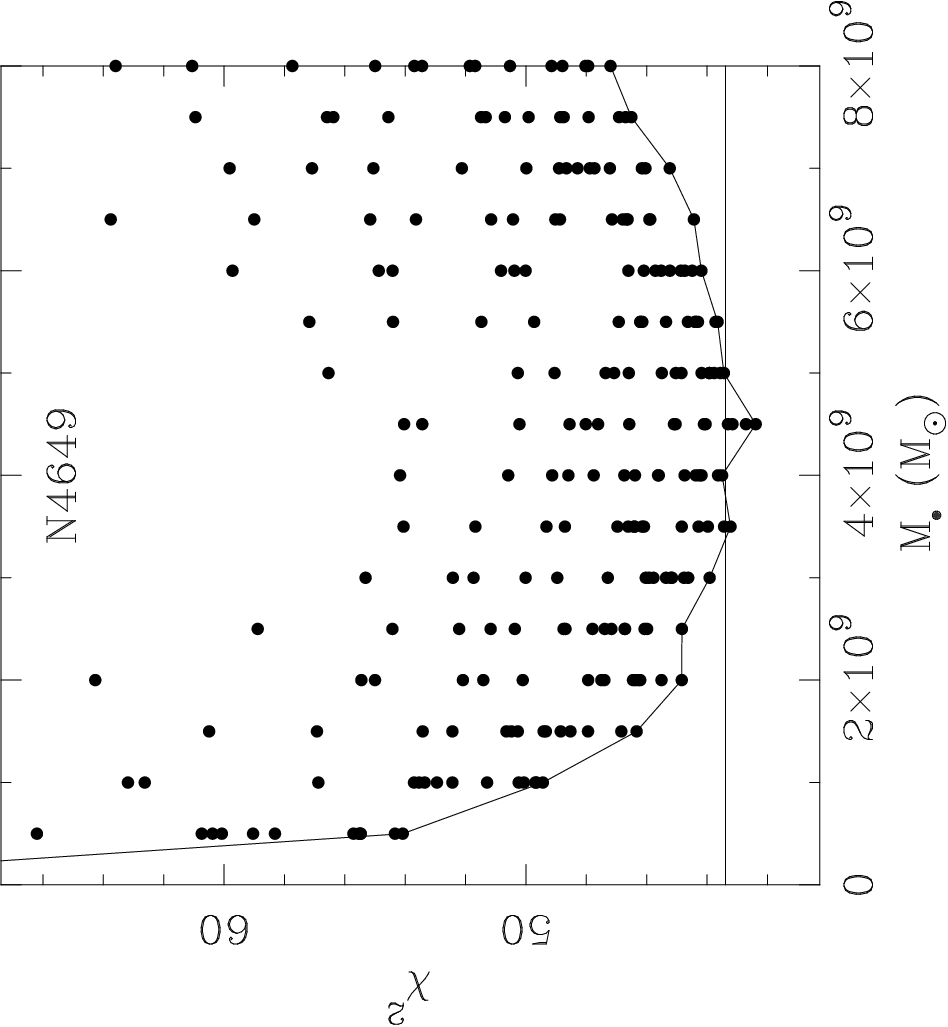}}
\vspace{0.02\hsize}
\centerline{
\includegraphics[angle=-90.,width=0.7\hsize]{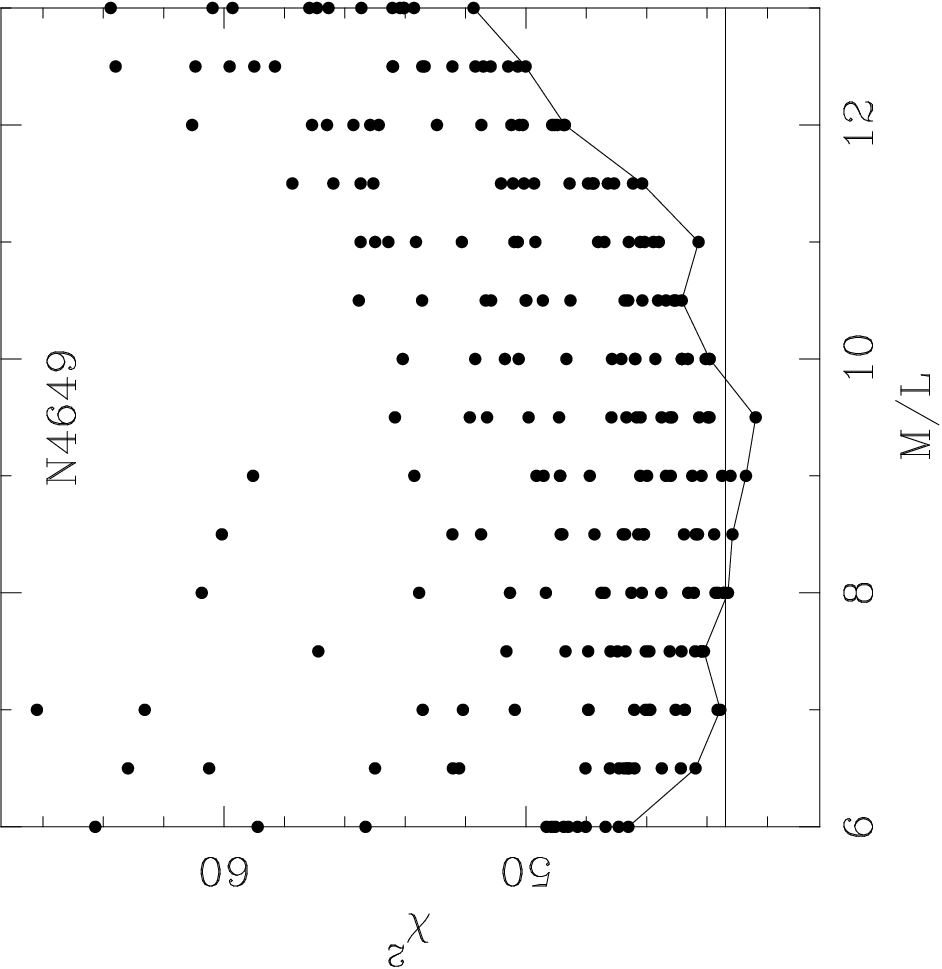}}
\vspace{0.02\hsize}
\centerline{
\includegraphics[angle=-90.,width=0.7\hsize]{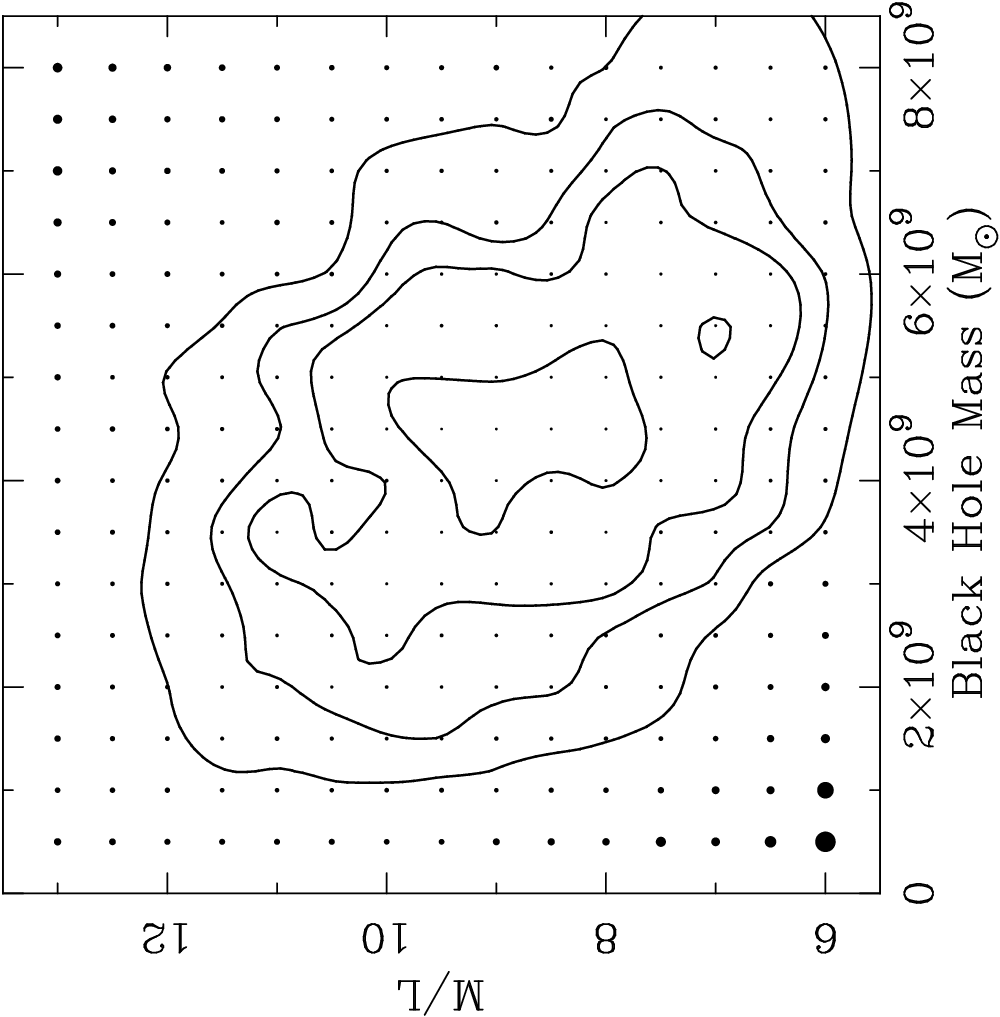}}

\caption{As in Figure~\ref{fig:xrayfit}, but for the $\chi^2$
distribution when no dark halo is included in the modeling.  Large
radii kinematics are excluded in the models. The absolute values of
$\chi^2$ in this figure are not comparable to those in
Figure~\ref{fig:chi2_4p} and \ref{fig:xrayfit}, as this figure fits to
less kinematic data (75 data points from the one LOSVD for the globular
kinematics and 4 LOSVDs for stellar kinematics beyond 40\arcsec are
excluded).}
\label{fig:noDM}
\end{figure}

\subsubsection{Comparison with the X-ray results}
\label{sec:comparexray}

In Figure~\ref{fig:dm} we also include the the mass profile inferred
from X-ray modelings \citep{gas_etal_07,hum_etal_08}, shown as the
green circles.  They are derived from the traditional approach, i.e.,
from the smoothed parametric fits to the temperature and density of
the X-ray emitting gas assuming hydrostatic equilibrium. The green
line in Figure~\ref{fig:dm} is our best representation of the X-ray
derived mass, assuming a logarithmic dark halo. The parameters for the
green line are $\mbh=3.5\times 10^9 \; \Msun$, $\mlvobs=5.0$, $\Vc=410
\kms$, and $\Rc =10 \kpc$. In Figure~\ref{fig:ML_r} the green solid
and dotted lines are the $\mlvobs$ of our best representation of the
X-ray potential without and with a black hole, respectively.

The total mass profile obtained from our modeling is consistently
larger than that from X-rays over most of the radial range by a factor
of about 1.7. Our stellar mass alone is larger than the X-ray derived
mass, so the difference cannot be due to the different dark matter
halos derived from the two methods. Our best-fit black hole mass is
consistent within the 1$\sigma$ error, yet about 35\% larger than, of
the $\mbh=3.35\times 10^9 \Msun$ derived in
\citet{hum_etal_08}. 
Detailed discussions on the difference from the X-ray mass profile are
in \S\ref{sec:discussmass}.

A concern from this work could be that the dark halo parameters are
poorly constrained by the globular cluster kinematics, leading to a
biased result on the mass profile and potentially on the black hole
mass. We therefore include the X-ray data in two different ways to
compare with our dynamical models. First, we use the X-rays to
determine the large radii mass profile, allowing our dynamical models
to determine the stellar M/L and black hole mass. Second, we adopt the
full X-ray mass profile as measured by \citet{hum_etal_08} and compare
with our best-fit model.

First, to use the large radii mass profile as determined from the
X-rays, we run a grid of models assuming the X-ray dark halo potential
($\Vc=410 \kms$, and $\Rc =10 \kpc$), and fit for black hole mass and
stellar $\mlvobs$. In other words, we fix only the outer part of the
green line in Figure~\ref{fig:dm}, and allow the inner mass profile to
seek the best fit to the data. The results are shown in the
distribution of $\chi^2$ as function of $\mbh$ and $\mlvobs$ in
Figure~\ref{fig:xrayfit}. We still find very similar $\mbh$ and
$\mlvobs$ as in the canonical set of models (\S\ref{sec:canonical});
both the best-fit values and the $1\sigma$ uncertainties of $\mbh$ and
$\mlvobs$ in Figure~\ref{fig:xrayfit} are very similar to those found
in Figure~\ref{fig:chi2_4p} and Figure~\ref{fig:corr}.

Second, we run the mass model with the X-ray derived parameters
($\mbh=3.5\times 10^9 \; \Msun$, $\mlvobs=5.0$, $\Vc=410 \kms$, and
$\Rc =10 \kpc$), and find the best fit to the kinematic data. The
$\chi^2$ of this model is about $98$, which is a much worse fit than
our best-fit model ($\chi^2\approx 43.5$). So the stellar and globular
cluster kinematics are indeed inconsistent with the X-ray derived mass
profile. Figure~\ref{fig:dchi2_r} demonstrates how $\Delta \chi^2$
changes with radius, where $\Delta \chi^2$ is the difference in the
radius-cumulated $\chi^2$ between the best-fit model in
Figure~\ref{fig:chi2_4p} ($\mbh=4.5\times 10^9 \; \Msun$,
$\mlvobs=9.0$, $\Vc=550 \kms$, and $\Rc =15 \kpc$) and the model with
the X-ray derived parameters (the green line in Figure~\ref{fig:dm},
$\mbh=3.5\times 10^9 \; \Msun$, $\mlvobs=5.0$, $\Vc=410 \kms$, and
$\Rc =10 \kpc$). The roughly linear increase of $\Delta \chi^2$ with
radius indicates that data from all radial range contribute equally to
the difference in the total $\chi^2$. This is another reason why the
assumed outer potential from X-rays does not affect the best-fit
$\mbh$ and $\mlvobs$.

\subsection{Models without a dark matter halo}

\citet{geb_tho_09} found that including a proper dark matter halo in
the dynamical modeling of M87 can give a factor of two larger black
hole mass than excluding one. Here we also do a similar test on how
the black hole mass and $\mlvobs$ change if we exclude a dark halo in
the modeling of NGC~4649 (i.e., the mass distribution is composed of
the black hole and the stars only, $\rho = \mbh \delta(r) +
\Upsilon\nu$).  Since we know beforehand that the dark halo dominates
at very large radii, we intentionally exclude some of the large radii
kinematics in order to not be unduly influenced by data that extend
well into the dark halo regime. The excluded regions are the one LOSVD
for the globular cluster kinematics and the four LOSVDs for the
stellar kinematics beyond 40\arcsec. We test the effect of including
all data, without a dark halo, and find that the resultant stellar
$\mlvobs$ (a constant independent of radius) is biased to a large
value of 11.5, and the minimal $\chi^2$ is larger by $\sim25$. Thus,
the large radii kinematics are excluded in the no-halo modeling.
Figure~\ref{fig:noDM} presents the $\chi^2$ distribution when a dark
halo is excluded and the large radii kinematics are excluded. From the
marginalized $\chi^2$ as a function of $\mbh$ and $\mlvobs$ in
Figure~\ref{fig:noDM}, we find the black hole mass $\mbh = (4.3\pm
0.7) \times 10^9\Msun$ and the $V$-band stellar $\mlvobs = 9.0 \pm
1.0$. The uncertainties with the no dark halo models are slightly
lower than when including a dark halo, and this difference could be
due to noise in the $\chi^2$ contours in Figure~\ref{fig:noDM}. The
best-fit black hole mass is quite different from the previously
published ones for NGC~4649 in \citet{geb_etal_03}, which are cruder
models without a dark halo either; we will discuss this point in
detail in \S\ref{sec:discussions}. Also these values differ from those
when a dark matter halo is allowed (\S~\ref{sec:canonical}) by less
than 5\%. So in NGC~4649 the inclusion of a proper dark halo does not
make as big a difference to the best-fit $\mbh$ and $\mlvobs$ as in
the case of M87. The existence of {\it HST} data in NGC~4649 certainly
helps to alleviate the bias. Also the globular cluster kinematics in
NGC~4649 are sparse (only one radial bin), so they do not have as much
leverage in the total $\chi^2$ as in M87. Furthermore, the dark matter
halo of NGC~4649 is not as massive and concentrated as that of M87, so
the black hole mass estimate of NGC~4649 is not biased as much by
neglecting the dark matter halo. 

We further experimented recomputing for the best-fit model using the
ground-based spectral data only, excluding the {\it HST} spectra. The
inclusion of the {\it HST} spectra in the modeling undoubtedly
improves the significance of the black hole substantially; the
$\chi^2$ difference between models with the best-fit $\mbh$ and with
zero $\mbh$ is around 30. The best-fit black hole mass is almost the
same, however, the uncertainty in $\mbh$ becomes much larger when the
{\it HST} spectra are excluded. This experiment emphasizes the need
for the {\it HST} spectra to improve the error in the determination of
$\mbh$.

\subsection{Velocity Dispersion Tensor}
\label{sec:disp}

\begin{figure}[!ht]
\centerline{
\includegraphics[angle=0.,width=0.9\hsize]{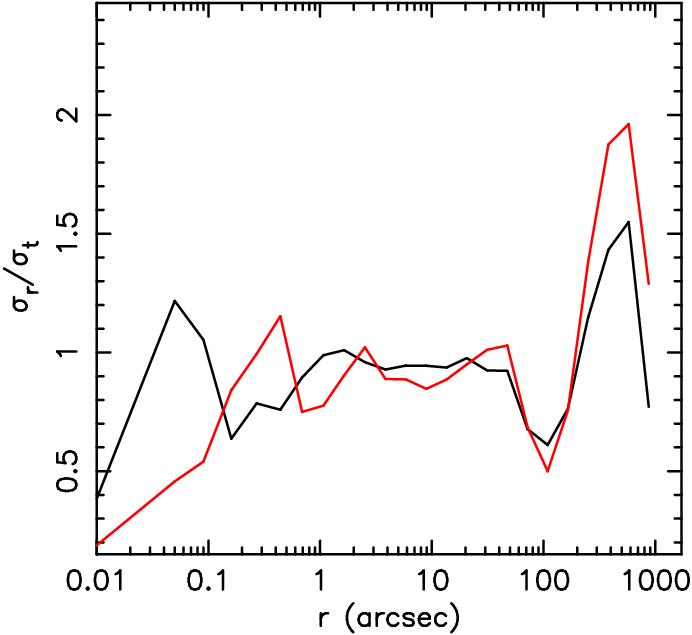}}

\caption{Shape of the velocity dispersion tensor as a function of radius for the best-fit model.  The black line is along the major axis, and the red line is near the minor axis.}
\label{fig:vr_vtang}
\end{figure}

\begin{figure}[!ht]
\centerline{
\includegraphics[angle=0.,width=0.9\hsize]{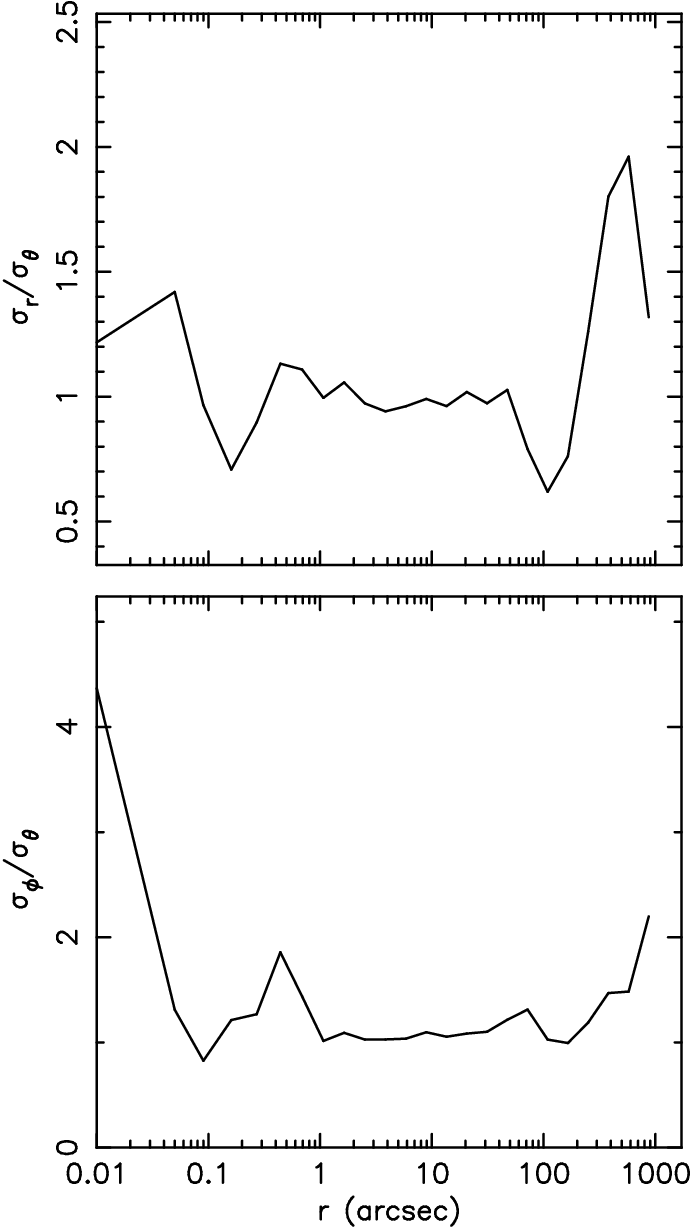}}
\caption{Ratios of the velocity dispersions $\sigmar$/$\sigmat$ (top) and $\sigmap$/$\sigmat$ (bottom) as a function of radius near the equatorial plane for the best-fit model.}
\label{fig:vr_vtheta}
\end{figure}

We also examine the internal orbital structures by studying the shape
of the velocity dispersion tensor. In our work the tangential
dispersion is defined as $\sigmatang=\sqrt{\sigmap^2 + \sigmat^2}$,
where $\sigmap$ includes contributions from both random and ordered
motions, i.e., it is the second moment of the azimuthal velocity
relative to the systemic velocity instead of to the mean rotation
velocity. An non-rotating isotropic model gives
$\sigmar$/$\sigmatang=1$. Figure~\ref{fig:vr_vtang} shows the internal
dispersion ratio $\sigmar$/$\sigmatang$ along the major and minor axes
for our best fit model ($\mbh=4.5\times 10^9 \; \Msun$, $\mlvobs=9.0$,
$\Vc=550 \kms$, and $\Rc =15 \kpc$). Over most of the radial range,
the model is very isotropic. It is strongly tangentially biased near
the center, and more radially biased at large
radii. \citet{geb_etal_03} discussed the possibility that the amount
of the tangential anisotropy at large radii may be overestimated when
the dark matter halo is excluded. The current result shows the orbits
are clearly more radially dominated at $r>100\arcsec$ when the dark
matter starts to dominate.

We also want to inspect if the distribution function of NGC~4649 truly
depends on three integrals of motion or only two: energy $E$ and the
$z$-component of the angular momentum $L_z$. We know that in a
two-integral model, $f=f(E, L_z)=f(v_R^2/2+
v_\phi^2/2+v_z^2/2+\Phi(R,z), Rv_\phi)$, $\sigma_R$ must be equal to
$\sigma_z$. One way to test whether a two-integral model is sufficient
is by comparing $\sigmar$ and $\sigmat$ on the equatorial plane where
$\sigma_R=\sigma_r|_{\phi=0}$ and $\sigma_z=\sigmat|_{\phi=0}$
(Figure~\ref{fig:vr_vtheta}). The radial variation of
$\sigmar/\sigmat$ in Figure~\ref{fig:vr_vtheta} demonstrates that the
best-fit model is inconsistent with having only two integrals of
motion, despite that the radially-averaged $\sigmar/\sigmat$ is close
to one. Thus, our three-integral model is needed to best fit the
data. We also show the $\sigmap/\sigmat$ ratio in the bottom panel of
Figure~\ref{fig:vr_vtheta}. We can see that near the black hole the
tangential dispersion is dominated by the azimuthal $\phi$ motion,
whereas the $\theta$ and $\phi$ dispersions are similar at most of
radii. Such detailed information on the shape of velocity dispersion
tensor could be potential constraints on the formation processes of
the supermassive black holes.


\section{Discussions}
\label{sec:discussions}

\subsection{Estimates of the mass profile}
\label{sec:discussmass}

From \S\ref{sec:comparexray} we see that the total mass profile from
our modeling is consistently larger than that from X-rays over most of
the radial range by a factor of about 1.7. Also the slopes of the two
mass profiles inside the effective radius are quite similar.  If our
orbit-based modeling results are confirmed in other independent
studies, then it may suggest that the X-ray derived mass is
systematically lower. There are several possible reasons why the X-ray
modeling could underestimate the true mass profile.

X-ray modelings generally assume that the X-ray emitting hot gas is in
hydrostatic equilibrium.  \citet{die_sta_07} argue that the X-ray gas
may not be in good hydrostatic equilibrium because they found no
correlation between X-ray and optical ellipticities in the inner
region where stellar mass dominates over dark matter and a shallow
correlation should be expected. If the hot gas is not in hydrostatic
equilibrium, the inflow of gas can make the X-ray derived mass smaller
than the true value \citep{pel_cio_06,joh_etal_09}.

Even if the hydrostatic equilibrium holds reasonably well for the
X-ray emitting gas, the existence of non-thermal pressure components,
but omitted in the X-ray modeling, can still make the X-ray derived
mass estimate less than the true value. For example, possible forms of
non-thermal pressure support include magnetic field, cosmic rays, and
microturbulence
\citep[e.g.][]{bri_mat_97,chu_etal_08,joh_etal_09}. Also, since the
hot gas in ellipticals comes mainly from the stellar mass loss, infall
or mergers, the gas could carry significant amount of angular
momentum. The gas can settle into rotationally supported systems as it
flows in and spins up. The spherically-determined X-ray mass may again
be an underestimate since the rotation is ignored.

However, \citet{bri_etal_09} study NGC~4649 and conclude that
turbulant motion in the X-ray gas is not strong enough to bias the
derived potential. Our dynamical results are in disagreement with the
X-ray derived potential, possibly suggesting that other non-thermal
pressures are present. While it is still unclear what combination of
systematics contribute to the difference in the mass estimates of the
X-ray and orbit-based methods, it is imperative to extend our
orbit-based modeling to a larger sample of galaxies that have been
analyzed with X-rays, and to examine if the difference between the two
methods is general.

\subsection{Black Hole Mass}

\begin{figure}[!ht]
\centerline{
\includegraphics[angle=-90.,width=\hsize]{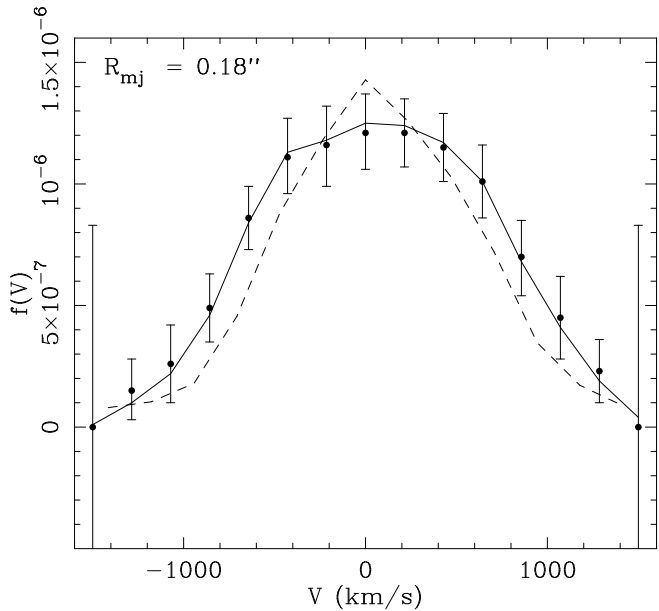}}
\caption{The modeled and observed LOSVDs at $R=0.18\arcsec$ on the
major axis. The y-axis represents the relative projected light for
this spatial bin. The data \citep{pin_etal_03} are shown by the solid
dots and their 68\% confidence bands (the velocity bins at the edge
have a large uncertainty because the original LOSVD extractions from
\citealt{pin_etal_03} do not extend that far in velocity space). The
solid line is the LOSVD from the best-fit model in this paper, and the
dashed line is the LOSVD from the best-fit model in
\citet{geb_etal_03}. The current model is a significantly improved fit
due to the inclusion of the highest angular momentum orbits (i.e.,
nearly circular orbits) that can match better the flat-topped LOSVD.}
\label{fig:compare_losvd}
\end{figure}

The black hole mass reported here is $4.5(\pm1.0)\times10^9\;
\Msun$. Using nearly the same input data, \citet{geb_etal_03} report a
black hole mass of $2.0(\pm0.5)\times10^9\; \Msun$ (note the reported
black hole mass of \citealt{geb_etal_03} has two changes applied to
it: first, we decrease the reported mass by 7\% for the different
distances, and second, we apply a 9\% increase in the mass due to a
numerical error in the previous calculation as reported in
\citealt{sio_etal_09}). While this is only a 1.5-$\sigma$ difference,
since we are using the same data, we should get the same result. Thus,
any difference must be in the modeling code changes, the input
assumptions (e.g., dark halo mass) or a combination. We find that the
difference is due to a more complete orbit sampling, discussed below,
but we have also checked a variety of other possibilities.

Comparing Figure~\ref{fig:noDM} (no dark halo) with
Figure~\ref{fig:chi2_4p} (including a dark halo) shows that there is
little difference in the black hole mass when including a dark
halo. This is expected since the black hole's effect on the kinematics
is well resolved with the {\it HST} kinematics. Other effects, such as
the radial range used in the modeling (in the present paper we must
extend the galaxy model out to larger radii to include the dark halo),
appear to have an insignificant effect as well.

The main modifications used for this study is to increase the orbit
library substantially and to sample the phase space differently. For
this study, we ran two sets of models, one with 30,000 orbits and one
with 13,000 orbits. The results are nearly identical in terms of
$\Delta\chi^2$ (which is what we use to determine the values and
uncertainties), however the minimum $\chi^2$ is lower by about
30. \citet{geb_etal_03} use 7000 orbits, with a $\chi^2$ that is
larger by 80. Given that we do not see a difference in the current
models using different orbit numbers, we are confident that orbit
number is not a concern for NGC~4649.

We do, however, find that the change in orbit sampling is the the main
factor causing difference in the black hole mass of NGC~4649. The
difference in orbit sampling is presented in \citet{tho_etal_04},
where we now use a sampling based on the density in the meridional
plane whereas the previous version (used in \citealt{geb_etal_03})
relied on sampling along the zero-velocity curve. In particular, the
new sampling covers more completely the phase space occupied by the
highest angular momentum orbits for a given energy, like near-circular
orbits and shell orbits.

Core galaxies appear to have significant tangential orbital anisotropy
in their centers \citep{geb_etal_03}. Thus, if the orbital coverage
poorly sampled important regions of phase space occupied by tangential
orbits, this may bias the best-fit model.  Indeed, the orbital
anisotropy is strongly biased towards tangential motion in the central
regions (Section~\ref{sec:disp}), much more so than in the original
orbit-models of \citet{geb_etal_03}. In Figure~\ref{fig:compare_losvd}
we exemplify the comparison of LOSVDs from the old and new orbit
sampling, and the observed LOSVD. In the figure the $HST$ observed
LOSVD at $R=0.18\arcsec$ on the major axis is quite flat-topped. The
modeled LOSVD in the current best-fit model (solid curve) is able
to fit the data very well, whereas the best-fit model in
\citet{geb_etal_03} (dashed curve) has trouble matching the
profile. The poor fit from the previous model is due to the lack of
nearly circular orbits.

Increasing the amount of tangential anisotropy causes the projected
dispersion to drop in the central regions; thus, in order to match the
observed projected velocity dispersion, the black hole mass has to be
increased to compensate. This concern needs to be tested on other
galaxies, specifically core galaxies. We note that the work of
\citet{sio_etal_09} compare black hole masses from analytic models
using the same orbital sampling presented here. They find no bias in
the black hole masses, but the central density of their galaxy
(NGC~4258) is cuspy instead of cored. Galaxies with such a large core
as NGC~4649 and large tangential orbital bias should be further
studied.

With $\mbh= 4.5 \times 10^9\Msun$ and the effective stellar velocity
dispersion $\sigma_e=385\;\kms$, NGC~4649's position is above the
latest $\mbh-\sigma_e$ relation \citep{gul_etal_09} by about 0.33 dex,
but still within the intrinsic scatter of the relation. Increasing the
largest black hole masses by a factor of 2---NGC~4649 in this paper
and M87 in \citet{geb_tho_09}---will have important consequences for
understanding the upper mass end (as in \citealt{lau_etal_07}) and
comparison with masses as estimated for quasars. There has been a long
standing problem as to why some quasars have black hole mass estimates
approaching $10^{10}~\Msun$, whereas none have ever been measured that
high even though the volumes surveyed should be large to see these; a
factor of 2 to 3 increase appears to resolve this issue. Furthermore,
an underestimate of the black hole mass seriously effects the physical
correlations, used extensively to quantify the role of black holes on
galaxy evolution (e.g., \citealt{hop_etal_08}), the amount of
deviation from a simple $\mbh-\sigma_*$ power-law relation
\citep{wyithe_06}, and the number density of the largest black holes
\citep{lau_etal_07,ber_etal_07}.

\subsection{Main uncertainties}

The uncertainties presented in this paper are statistical
uncertainties from the $\chi^2$ analysis. In order to translate a
given change in $\chi^2$ into a significance (for example, we report
the 68\% confidence band based on $\Delta\chi^2=1$), we make two major
assumptions.

First, we assume that we understand the uncertainties of the
kinematics and any correlations between the measurements. Correlations
in the spectral extractions of the LOSVDs may cause biases in the
determination of the parameter uncertainties such as the black hole
mass (discussed in \citealt{hou_etal_06}). For the analysis used in
this paper, there are likely correlations due to smoothing of the
LOSVD, as discussed originally in
\citet{geb_etal_00}. \citet{geb_etal_04} tries to quantify this effect
by running Monte Carlo simulations, starting from the noise in the
spectra (and then running dynamical models for each realization). A
most complete analysis is in \citet{sio_etal_09} who find a similar
result (see their Fig. 17). The uncertainty as measured from the
$\Delta\chi^2$ analysis is similar to, but smaller than, the
uncertainty as measured from the spread in the best-fitted values from
the realizations. This work suggests that the $\chi^2$ analysis
provides realistic uncertainties. Yet, we know there are correlations
in the LOSVDs---these correlations exist in the non-parametric
analysis that was used in this paper but also exist in the basis
function approach advocated by \citet{hou_etal_06} due to degeneracies
with the continuum placement for the spectra. In this analysis of
NGC~4649, we rely on $\Delta\chi^2=1$ reflecting the 68\% confidence
band. From the tests we describe above, we argue that the statistical
analysis is understood, although expanded analysis along the lines
presented in \citet{geb_etal_04} and \citet{sio_etal_09} is greatly
desired. We feel the more immediate need is to control systematic
effects, discussed next.

Second, our reported uncertainties do not include effects from
systematic uncertainties. As is clear from the change in the black
hole mass from the previous study, there are systematic uncertainties
that need to be explored. These uncertainties will influence the black
hole mass, orbital structure, and dark halo profile. We discuss above
the concern with the orbital sampling. We note, again, that the orbit
sampling issue is likely only important for galaxies that have an
extreme orbital structure as seen here in
NGC~4649. \citet{geb_etal_04} tested the effect of orbit number on a
power-law galaxy and found no bias in the black hole mass
estimate. Running analytic models with a range of orbital
distributions would be worthwhile, as well as studying additional core
galaxies. For core galaxies, another important systematic is that they
may be better modeled as triaxial as opposed to axisymmetric as
assumed here (e.g., \citealt{vdb_etal_08}). \citet{vdb_dez_09} show that the
black hole mass in the core galaxy NGC~3379 increases if the model is
allowed to be triaxial. This increase could be particular to the
intrinsic orientation for NGC~3379 but continued studies including
triaxiality is warranted. Other model assumptions such as the assumed
inclination, variation of the stellar mass-to-light ratio with radius
(as derived from stellar population models), and the assumed shape of
the dark halo profile, for example, could be additional sources of
systematic uncertainties.


\section{Conclusions}
\label{sec:conclusions}

We model the dynamical structure of NGC~4649 using the high resolution
data sets from {\it HST}, stellar, and globular cluster
observations. Our main new results are:

1. Our modeling gives $\mbh= 4.5 \pm 1.0 \times 10^9\Msun$ and
   $\mlvobs=8.7 \pm 1.0$.  Our new $\mbh$ of NGC~4649 is about a
   factor of 2 larger than the previous result. We find that the
   earlier model did not adequately sample the orbits required to
   match the large tangential anisotropy in the galaxy center..

2. We confirm the presence of a dark matter halo in NGC~4649, but the
   stellar mass dominates inside the effective radius. The parameters
   of the dark halo especially the core radius are less constrained
   due to the sparse globular cluster data at large radii.

3. Unlike in the case of M87, the black hole mass from the dynamical
   modeling is not biased as much by the inclusion of a dark matter
   halo, because high-resolution {\it HST} spectra are available for
   NGC~4649, the globular cluster kinematics are sparse, and the halo
   is not as dominant inside the effective radius $R_e$ as that of
   M87.

4. We find that in NGC~4649 the dynamical mass profile from our
   modeling is consistently larger than that derived from the X-ray
   data over most of the radial range by about 70\%. It implies that
   either some forms of non-thermal pressure need to be included, the
   assumed hydrostatic equilibrium may not be a good approximation in
   the X-ray modeling, or the assumptions used in our dynamical
   modeling create a bias.

\acknowledgements We thank Remco van den Bosch for helpful comments on
the manuscript. This work was supported by NSF-CAREER grant
AST03-49095. J.S. acknowledges partial support from a Harlan J. Smith
fellowship at the McDonald Observatory of the University of Texas at
Austin. We also acknowledge the generous support from the Texas
Advanced Computing Center with their state-of-art computing
facilities.


\newcommand{\noopsort}[1]{} \newcommand{\singleletter}[1]{#1}

\newpage

\end{document}